%% file: ms.tex
\documentclass[letterpaper, 10 pt, conference]{ieeeconf} 
\IEEEoverridecommandlockouts   

\makeatletter
\def\endthebibliography{%
  \def\@noitemerr{\@latex@warning{Empty `thebibliography' environment}}%
  \endlist
}
\def\input@path{{./Figures}}
\makeatother

\usepackage{cite}
\usepackage{amsmath,amssymb,amsfonts}
\usepackage{algorithmic}
\usepackage{textcomp}
\usepackage{xcolor}
\usepackage[normalem]{ulem}

\def\BibTeX{{\rm B\kern-.05em{\sc i\kern-.025em b}\kern-.08em T\kern-.1667em\lower.7ex\hbox{E}\kern-.125emX}}

\usepackage{amsthm}
\newtheoremstyle{dotless}{}{}{\itshape}{}{\bfseries}{}{ }{}

\let\labelindent\relax 
\usepackage{enumitem}

\usepackage{graphicx}
\usepackage{pgfplots}
\usepackage{ifthen}
\usepackage{eso-pic}

\usetikzlibrary{plotmarks}
\usetikzlibrary{arrows.meta}
\usetikzlibrary{patterns,ipe}
\usepgfplotslibrary{patchplots}

\theoremstyle{dotless}
\newtheorem{dummy}{}
\newtheorem{assumption}[dummy]{Assumption}
\newtheorem{theorem}[dummy]{Theorem}
\newtheorem{corollary}[dummy]{Corollary}
\newtheorem{lemma}[dummy]{Lemma}
\newtheorem{procedure}[dummy]{Algorithm}

\newtheorem{definition}[dummy]{Definition}
\newtheorem{remark}[dummy]{Remark}

\newcommand{\norm}[2]{\lVert#1\rVert_{#2}}

\newcommand{\im}{\textnormal{im}}

\newcommand{\Rq}{\mathbb{R}[q^{-1}]}
\newcommand{\T}{T}
\newcommand{\fhl}{\underline{\hat{f}}}
\newcommand{\rl}{\underline{r}}

\newcommand{\drawlinelegend}[1]{\raisebox{.5ex}{\tikz{\draw[#1, line width=0.4mm] (0,0) -- +(1em, 0);}}}

\newboolean{shorten}
\setboolean{shorten}{false}
\renewcommand{\baselinestretch}{0.97} 
\makeatletter
\g@addto@macro\normalsize{%
}
\makeatother

\AddToShipoutPictureBG*{%
	\AtPageUpperLeft{%
		\setlength\unitlength{1in}%
		\hspace*{\dimexpr0.5\paperwidth\relax}
		\makebox(0,-0.75)[c]{\begin{tabular}{c c}
				Johan Kon, Unifying Model-Based and Neural Network Feedforward: \\ Physics-Guided Neural Networks with Linear Autoregressive Dynamics, \\
				To appear in {\em 2022 IEEE Conference on Decision and Control}, Cancún, Mexico, 2022, uploaded to ArXiv \today
		\end{tabular}}
}}

\begin{document}

\bstctlcite{IEEEexample:BSTcontrol} 



\title{\LARGE \bf Unifying Model-Based and Neural Network Feedforward: \\ Physics-Guided Neural Networks with Linear Autoregressive Dynamics 
}

\author{Johan Kon$^1$, Dennis Bruijnen$^2$, Jeroen van de Wijdeven$^3$, Marcel Heertjes$^{1,3}$, and Tom Oomen$^{1,4}$
\thanks{This work is supported by Topconsortia voor Kennis en Innovatie (TKI), and ASML and Philips Engineering Solutions. $^1$: Control Systems Technology Group, Departement of Mechanical Engineering, Eindhoven University of Technology, P.O. Box 513, 5600 MB Eindhoven, The Netherlands, e-mail: j.j.kon@tue.nl. $^2$: Philips Engineering Solutions, High Tech Campus 34, 5656 AE Eindhoven, The Netherlands. $^3$: ASML, De Run 6501, 5504 DR Veldhoven, The Netherlands. $^4$: Delft University of Technology, P.O. Box 5, 2600 AA Delft, The Netherlands. }
}

\maketitle

\input{Sections/Abstract.tex}

\input{Sections/Introduction.tex}

\input{Sections/Problem_formulation.tex}

\input{Sections/SK_iterations.tex}

\input{Sections/non_uniqueness_EE_LS.tex}

\input{Sections/SK_and_Orthogonal_regularization.tex}

\input{Sections/Simulation_validation.tex}

\input{Sections/Conclusion}

\bibliographystyle{IEEEtran}
\bibliography{IEEEabrv,library.bib}

\input{Sections/Appendix}

\end{document}

%% file: Sections/Abstract.tex
\begin{abstract}
Unknown nonlinear dynamics often limit the tracking performance of feedforward control.
The aim of this paper is to develop a feedforward control framework that can compensate these unknown nonlinear dynamics using universal function approximators.
The feedforward controller is parametrized as a parallel combination of a physics-based model and a neural network, where both share the same linear autoregressive (AR) dynamics. 
This parametrization allows for efficient output-error optimization through Sanathanan-Koerner (SK) iterations. Within each SK-iteration, the output of the neural network is penalized in the subspace of the physics-based model through orthogonal projection-based regularization, such that the neural network captures only the unmodelled dynamics, resulting in interpretable models. 

\end{abstract}

%% file: Sections/Introduction.tex
\section{Introduction}
\label{sec:Introduction}
Feedforward control can significantly increase the performance of dynamic systems \cite{Clayton2009, 489285}, e.g., positioning accuracy in motion systems. In feedforward control, the key requirements are high tracking performance and task flexibility \cite{Butterworth2009}, i.e., a small tracking error for a variety of references. Additionally, it is often desired that the feedforward controller is interpretable \cite{Schoukens2019}, and that its parameters can be efficiently learned given a training dataset. 
%
%

Feedforward controllers based on physical models are highly flexible and interpretable by design \cite{Lambrechts2005}. For example, the dynamics can be parametrized as a rational transfer function \cite{Zou2009} \ifthenelse{\boolean{shorten}}{\textcolor{red}{\sout{representing a lumped-mass model.}}}{}
These parametrizations allow for efficient optimization \cite{Sanathanan1963} and can be interpreted through frequency-domain tools, e.g., Bode diagrams. Extensions include static friction \cite{Boerlage2003} and position-varying compliance feedforward \cite{Kontaras2017}, as well as methods to compensate for nonminimum-phase zero dynamics \cite{Devasia1996}. However, these physics-based parametrizations often have limited performance in the presence of unknown, typically nonlinear dynamics \cite{6837472, Ljung2020}.

On the other hand, feedforward signals that compensate all reproducible dynamics, i.e., achieve tracking performance up to the noise level of the system, can be generated through learning control methods such as iterative learning control (ILC) \cite{Bristow2006}. Yet, these approaches lack task flexibility, necessitating the use of, e.g., basis functions \cite{6837472}, and do not result in interpretable feedforward signals.

To go beyond the trade-off between performance and task flexibility, universal function approximators \textcolor{red}{ \ifthenelse{\boolean{shorten}}{\sout{from the field of machine learning}}{}}such as neural networks have been used as flexible feedforward parametrizations \cite{HUNT19921083},\textcolor{red}{\ifthenelse{\boolean{shorten}}{\sout{consequently}}{}} overcoming the performance decrease of physics-based parametrizations in the context of unmodelled dynamics. Examples include nonlinear auto-regressive exogenous (NARX) and nonlinear finite impulse response (NFIR) parametrizations \cite{Sjoberg1995, 80202}, and long short-term memory neural networks \cite{Ljung2020}. As a downside, these parametrizations are not interpretable, and learning their parameters is computationally challenging. Additionally, these universal approximators lack the ability to extrapolate \cite{Schoukens2019}, deteriorating task flexibility outside the training regime.

Physics-guided neural networks (PGNNs) \cite{7959606, 2017arXiv171011431K} are a combined model-approximator parametrization and aim to reconcile the interpretability and task flexibility of model-based approaches with the performance of universal function approximators. Physics-guided parametrizations indeed significantly improve performance over model-based feedforward alternatives \cite{Bolderman2021}. Interpretability is obtained through explicitly separating the neural network and model contribution by imposing orthogonality  \cite{Kon2022PhysicsGuided}. Even so, the performance of these PGNNs is limited\textcolor{red}{ \ifthenelse{\boolean{shorten}}{ \sout{since they cannot capture zero dynamics or compensate for flexible modes.}}{} }as they do not contain AR dynamics and thus cannot compensate for zero dynamics of the system.

Although major steps have been taken to improve the flexibility of data-driven feedforward control while maintaining interpretability, at present these are limited by existing classes of PGNNs that can only handle overly simplified system dynamics.
The aim of this paper, therefore, is to develop a class of PGNNs for feedforward control that can compensate zero dynamics\textcolor{red}{\ifthenelse{\boolean{shorten}}{ \sout{resulting from flexible modes}}{}}.
The main contribution is a \textcolor{red}{\ifthenelse{\boolean{shorten}}{\sout{feedforward control framework consisting of a parallel combination of a physics-based model for task flexibility, and a neural network that learns only unmodelled dynamics with the aim of increasing performance.}}{}}PGNN feedforward control framework with AR dynamics, in which the model is interpretable and the neural network learns only unmodelled dynamics.
This is achieved through the following subcontributions:
\begin{enumerate}[label=C\arabic*)]
	\vspace{-2pt}
	\item A physics-guided feedforward parametrization with shared linear autoregressive dynamics (Section \ref{sec:problem_formulation}).
	\item An efficient output-error optimization algorithm based on SK-iterations \cite{Sanathanan1963} (Section \ref{sec:SK_iterations} and \ref{sec:SK_and_orthogonal_reg}).
	\item An orthogonal projection-based regularizer promoting orthogonality of the model and neural network, ensuring interpretability of the model (Section \ref{sec:OP_regularizer}).
	\vspace{-1pt}
\end{enumerate}
%
\subsubsection*{Notation and Definitions}
All systems are discrete-time with sample time $T_s$. The sets $\mathbb{Z}_{> 0}$, $\mathbb{R}_{\geq 0}$ represent the set of positive integers and non-negative real numbers. For the signal $u$ with length $N$, $u(k) \in \mathbb{R}$ represents the signal at time index $k = \mathbb{Z}_{[1,N]}$, whereas $\underline{u} = \begin{bmatrix}u(1) & \ldots & u(N) \end{bmatrix}^\T \in \mathbb{R}^{N}$ is its finite-time vector representation. The set $\Rq$ is the set of polynomials in $q^{-1}$ with real coefficients, with $q^{-1} u(k) = u(k-1)$. $\textrm{Id}(\cdot)$ represents the identity operator.

%% file: Sections/Problem_formulation.tex
\section{Problem Formulation}
\label{sec:problem_formulation}
In this section, first the problem of feedforward control for dynamic systems is introduced. Second, the physics-guided feedforward parametrization consisting of a physics-based model and neural network with shared linear AR dynamics is defined. Lastly, the learning problem is formulated.

\subsection{Feedforward Setup and Physics-Guided Parametrization}
\begin{figure}
\centering
\input{FFW_setup.tex}
\caption{Feedforward setup with input $f$, dynamic system $\mathcal{J}$, reference $r$, and error $e$ (left). The input $f$ is parametrized as the output of a reference dependent filter $\mathcal{F}_{\theta,\phi}$ (right).}
\label{fig:FFW_setup}
\end{figure}
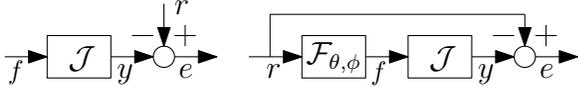

The goal of feedforward control, see Fig. \ref{fig:FFW_setup}, is to generate input $f(k) \in \mathbb{R}$ to the discrete-time system $\mathcal{J}$ such that its output $y(k) \in \mathbb{R}$ equals the desired output $r(k) \in \mathbb{R}$, i.e.,
\begin{equation}
	e(k) = r(k) - y(k) = r(k) - \mathcal{J}(f(k)) = 0 \quad \forall k \in \mathbb{Z}_{>0},
\end{equation}
with $e(k) \in \mathbb{R}$ the tracking error. The system $\mathcal{J}$ can represent a feedback-controlled or open-loop system. 
 
To obtain both high performance and task flexibility, the input signal $f$ is parametrized as the output of a feedforward controller acting on reference $r$. More specifically, the feedforward controller $\mathcal{F}_{\theta,\phi}$ is a parallel combination of a physics-based model $\mathcal{M}_\theta$ that is linear in its parameters (LIP) $\theta$, and universal approximator $\mathcal{C}_\phi$\textcolor{red}{\ifthenelse{\boolean{shorten}}{\sout{ ,which are defined next.}}{} }with parameters $\phi$.

\begin{definition}[Model class]
The model $\mathcal{M}_\theta: r(k) \rightarrow f_\mathcal{M}(k)$ satisfies the ordinary difference equation
\begin{align}
	f_\mathcal{M}(k) + \sum_{i=1}^{N_b} b_i q^{-i} f_\mathcal{M}(k) = \sum_{i=1}^{N_a} a_i g_i\left(\psi_i(r(k))\right),
	\label{eq:model_class}
\end{align}
with $\psi_i \in \Rq$ and static nonlinearities $g_i: \mathbb{R} \rightarrow \mathbb{R}$, both user-defined functions, and parameters $\theta = \{a_i\}_{i=1}^{N_a} \cup \{b_i\}_{i=1}^{N_b}$, $a_i,b_i \in \mathbb{R}$.
\end{definition}
\textcolor{red}{\ifthenelse{\boolean{shorten}}{\sout{Examples that fit in this model class are the class of rational transfer functions for $g_i = \textrm{Id}$ and $\psi_i = q^{-i}$. 
Additionally, nonlinear friction models such as Coulomb friction can be encapsulated as $g_i(\psi_i(r(k)) = \textrm{sign}(\delta r(k))$ with $\delta = \textcolor{red}{T_s^{-1}}(1 - q^{-1}) \in \Rq$ the discrete time derivative. Lastly, trigonometric nonlinearities that often result from first-principles modelling can be encapsulated as, e.g., for an inverse pendulum, $a_i g_i(\psi_i(r(k)) = m g l \cos(\phi)$.}}{} }Examples that can be encapsulated by this model class are, i.a., the class of rational transfer functions for $g_i = \textrm{Id}$ and $\psi_i = q^{-i}$, and trigonometric nonlinearities resulting first-principles modelling, such as $a_i g_i(\psi_i(r(k)) = m g l \cos(\phi)$ for an inverse pendulum.

\begin{figure}
\centering
\input{Feedforward_parametrization2.tex}
\caption{Feedforward filter $\mathcal{F}_{\theta,\phi}$ as the parallel combination of model $\mathcal{M}_\theta$ and approximator $\mathcal{C}_\phi$ sharing AR dynamics $B(q^{-1}) = 1 + \sum_{i=1}^{N_b} b_i q^{-i}$, in this example with 2 hidden layers of 3 neurons and no skip connections.}
\label{fig:FFW_filter}
\end{figure}
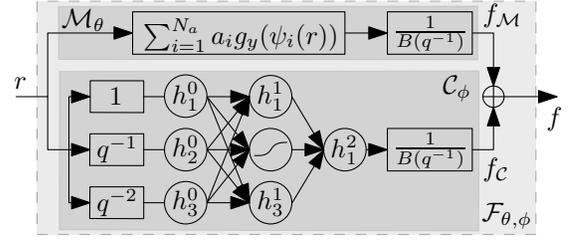

\begin{definition}[Approximator class]
	The approximator $\mathcal{C}_\phi: r(k) \rightarrow f_\mathcal{C}(k)$ satisfies the ordinary difference equation
	\begin{equation}
		f_\mathcal{C}(k) + \sum_{i=1}^{N_b} b_i q^{-i} f_\mathcal{C}(k) = g_{\phi}(r(k)),
	\end{equation}
	where $g_{\phi}(r(k))$ is the output of a neural network given by
\begin{align}
	h^l(r(k)) &= \begin{bmatrix} r(k), \ldots, r(k-q)\end{bmatrix}^T	 & \textrm{if}\ \ & l = 0 \nonumber \\
	h^l(r(k)) &= \sigma \left( W^{l-1} h^{l-1}(k) + c^l \right) & \textrm{if}\ \  & l = {1, \ldots, L} \nonumber \\
	g_{\phi}(r(k)) &= W^l h^{l}(r(k)) & \textrm{if}\ \ & l = L, \label{eq:FNN}
\end{align}
with $W^l \in \mathbb{R}^{N_l \times N_{l-1}}$ the weights and $c^l \in \mathbb{R}^{N_l}$ the biases of layer $l$ with $n_l$ neurons, $\sigma(\cdotp)$ an element-wise activation function, and parameter set $\phi = \{W^l, c^l \}_{l=0}^{L-1} \cup \{W^L\}$.
\end{definition}
The network $g_\phi(\cdot)$ acts on a past window of references $r(k)$, and is here represented by a fully connected multilayer perceptron without skip connections, see $\mathcal{C}_\phi$ in Fig. \ref{fig:FFW_filter}. It can be replaced by any network with a directed acyclic graph structure\textcolor{red}{\ifthenelse{\boolean{shorten}}{\sout{, such as convolutional and }}{}}, e.g., residual neural networks \cite{He2016}, including user-defined input transformations and a bias in the final layer.

Since $\mathcal{M}_\theta$ and $\mathcal{C}_\phi$ share the same linear AR dynamics $f(k) + \sum_{i=1}^{N_b} b_i q^{-i} f(k)$, the parallel combination $\mathcal{F}_{\theta,\phi}$, see Fig. \ref{fig:FFW_filter}, also has these linear AR dynamics, as defined next. 


\begin{definition}
The feedforward controller $\mathcal{F}_{\theta,\phi}: r(k) \rightarrow f(k)$ is given by 
\begin{align}
	\mathcal{F}_{\theta,\phi}(r(k)) = \mathcal{M}_{\theta}(r(k)) + \mathcal{C}_{\phi}(r(k)),
\end{align}
such that it satisfies
\begin{align}
	\underbrace{(1 + \sum_{i=1}^{N_b} b_i q^{-i} )}_{B(q)} f(k) &= \underbrace{\sum_{i=1}^{N_a} a_i g_i(\psi_i(r(k)))}_{A(r(k))} + g_\phi(r(k)). \label{eq:ffw_parametrization}
\end{align}
\end{definition}
The parametrization $\mathcal{F}_{\theta,\phi}$ has nonlinear exogenous dynamics $A(r(k)) + g_\phi(r(k))$ and linear AR dynamics $B(q)f(k)$. Therefore, $\mathcal{F}_{\theta,\phi}$ is less complex than a NARX parametrization \cite{Ljung2020} with nonlinear AR dynamics, but it can capture a relevant class of physical systems with linear zero dynamics, as shown in Section \ref{sec:simulation_results}, which cannot be captured by NFIR \cite{Kon2022PhysicsGuided} or rational transfer function \cite{6837472} parametrizations. In addition, the linear AR dynamics allow for linear stability analysis and inversion tools \cite{Zou2009a}, and for efficient output-error (OE) minimization through SK-iterations \cite{Sanathanan1963}.
%
%
%
%
%

\textcolor{red}{}To learn parameters $\theta,\phi$, a dataset $\mathcal{D} = \{r(k), \hat{f}(k)\}_{k=1}^N$ is assumed to be available with reference $r(k)$ and the corresponding input $\hat{f}$, such that $r(k) = \mathcal{J}(\hat{f}(k))$. This input $\hat{f}$ can be obtained by, e.g., ILC.
\ifthenelse{\boolean{shorten}}{
\begin{assumption}
\textcolor{red}{\sout{	For system $\mathcal{J}$ and a desired output $r(k)$ with $k \in \mathbb{Z}_{[1,N]}$, there exists an input $\hat{f}(k)$, such that $\mathcal{J}(\hat{f}(k)) = r(k) $, and these input/output pairs are available as a dataset $\mathcal{D} = \{r(k), \hat{f}(k)\}_{k=1}^N$ obtained by, e.g., ILC.}} 
\end{assumption}
\textcolor{red}{\sout{This dataset is used to learn the parameters of a feedforward controller that generates $\hat{f}$ based on $r$.}}
}{}
\subsection{Problem Formulation}
The aim of this paper is to learn parameters $\theta,\phi$ of $\mathcal{F}_{\theta,\phi}$ in \eqref{eq:ffw_parametrization} based on dataset $\mathcal{D}$, such that $f(k) = \hat{f}(k)$, implying $e(k) = 0 \ \forall k \in \mathbb{Z}_{>0}$. This includes
\begin{enumerate}[label=\arabic*)]
	\item an output error (OE) criterion\textcolor{red}{\ifthenelse{\boolean{shorten}}{\sout{ for feedforward parametrization \eqref{eq:ffw_parametrization}}}{}} that can be efficiently solved through SK-iterations because of the shared linear AR dynamics $B(q)f$,
	\item regularizing this OE criterion with an orthogonal projection-based regularizer to promote unique coefficients $\theta$, resulting in interpretable models, and
	\item illustrating the approach on a two-mass-damper-spring system with Stribeck-like friction characteristics.
\end{enumerate}

%% file: FFW_setup.tex
\tikzstyle{ipe stylesheet} = [
  ipe import,
  even odd rule,
  line join=round,
  line cap=butt,
  ipe pen normal/.style={line width=0.4},
  ipe pen heavier/.style={line width=0.8},
  ipe pen fat/.style={line width=1.2},
  ipe pen ultrafat/.style={line width=2},
  ipe pen normal,
  ipe mark normal/.style={ipe mark scale=3},
  ipe mark large/.style={ipe mark scale=5},
  ipe mark small/.style={ipe mark scale=2},
  ipe mark tiny/.style={ipe mark scale=1.1},
  ipe mark normal,
  /pgf/arrow keys/.cd,
  ipe arrow normal/.style={scale=7},
  ipe arrow large/.style={scale=10},
  ipe arrow small/.style={scale=5},
  ipe arrow tiny/.style={scale=3},
  ipe arrow normal,
  /tikz/.cd,
  ipe arrows, 
  <->/.tip = ipe normal,
  ipe dash normal/.style={dash pattern=},
  ipe dash dotted/.style={dash pattern=on 1bp off 3bp},
  ipe dash dashed/.style={dash pattern=on 4bp off 4bp},
  ipe dash dash dotted/.style={dash pattern=on 4bp off 2bp on 1bp off 2bp},
  ipe dash dash dot dotted/.style={dash pattern=on 4bp off 2bp on 1bp off 2bp on 1bp off 2bp},
  ipe dash normal,
  ipe node/.append style={font=\normalsize},
  ipe stretch normal/.style={ipe node stretch=1},
  ipe stretch normal,
  ipe opacity 10/.style={opacity=0.1},
  ipe opacity 30/.style={opacity=0.3},
  ipe opacity 50/.style={opacity=0.5},
  ipe opacity 75/.style={opacity=0.75},
  ipe opacity opaque/.style={opacity=1},
  ipe opacity opaque,
]
\definecolor{red}{rgb}{1,0,0}
\definecolor{blue}{rgb}{0,0,1}
\definecolor{green}{rgb}{0,1,0}
\definecolor{yellow}{rgb}{1,1,0}
\definecolor{orange}{rgb}{1,0.647,0}
\definecolor{gold}{rgb}{1,0.843,0}
\definecolor{purple}{rgb}{0.627,0.125,0.941}
\definecolor{gray}{rgb}{0.745,0.745,0.745}
\definecolor{brown}{rgb}{0.647,0.165,0.165}
\definecolor{navy}{rgb}{0,0,0.502}
\definecolor{pink}{rgb}{1,0.753,0.796}
\definecolor{seagreen}{rgb}{0.18,0.545,0.341}
\definecolor{turquoise}{rgb}{0.251,0.878,0.816}
\definecolor{violet}{rgb}{0.933,0.51,0.933}
\definecolor{darkblue}{rgb}{0,0,0.545}
\definecolor{darkcyan}{rgb}{0,0.545,0.545}
\definecolor{darkgray}{rgb}{0.663,0.663,0.663}
\definecolor{darkgreen}{rgb}{0,0.392,0}
\definecolor{darkmagenta}{rgb}{0.545,0,0.545}
\definecolor{darkorange}{rgb}{1,0.549,0}
\definecolor{darkred}{rgb}{0.545,0,0}
\definecolor{lightblue}{rgb}{0.678,0.847,0.902}
\definecolor{lightcyan}{rgb}{0.878,1,1}
\definecolor{lightgray}{rgb}{0.827,0.827,0.827}
\definecolor{lightgreen}{rgb}{0.565,0.933,0.565}
\definecolor{lightyellow}{rgb}{1,1,0.878}
\definecolor{black}{rgb}{0,0,0}
\definecolor{white}{rgb}{1,1,1}
\begin{tikzpicture}[ipe stylesheet]
  \draw
    (64, 652) rectangle (88, 636);
  \node[ipe node, font=\large]
     at (72, 640) {$\mathcal{J}$};
  \draw
    (108, 644) circle[radius=4];
  \draw[->]
    (108, 664)
     -- (108, 648);
  \node[ipe node, font=\large]
     at (112, 660) {$r$};
  \draw[->]
    (88, 644)
     -- (104, 644);
  \node[ipe node, font=\large]
     at (90.288, 636.208) {$y$};
  \draw[->]
    (112, 644)
     -- (128, 644);
  \draw
    (96, 652)
     -- (104, 652);
  \draw
    (116, 648)
     -- (116, 656);
  \draw
    (112, 652)
     -- (120, 652);
  \node[ipe node, font=\large]
     at (113.644, 636.22) {$e$};
  \draw[->]
    (48, 644)
     -- (64, 644);
  \node[ipe node]
     at (49.207, 635.668) {$f$};
  \draw
    (200, 652) rectangle (224, 636);
  \node[ipe node, font=\large]
     at (208, 640) {$\mathcal{J}$};
  \draw
    (244, 644) circle[radius=4];
  \draw[->]
    (244, 660)
     -- (244, 648);
  \draw[->]
    (224, 644)
     -- (240, 644);
  \node[ipe node, font=\large]
     at (226.288, 636.208) {$y$};
  \draw[->]
    (248, 644)
     -- (264, 644);
  \draw
    (232, 652)
     -- (240, 652);
  \draw
    (252, 648)
     -- (252, 656);
  \draw
    (248, 652)
     -- (256, 652);
  \node[ipe node, font=\large]
     at (249.644, 636.22) {$e$};
  \draw[->]
    (184, 644)
     -- (200, 644);
  \node[ipe node]
     at (186.018, 635.397) {$f$};
  \draw
    (160, 652) rectangle (184, 636);
  \draw[->]
    (140, 644)
     -- (160, 644);
  \node[ipe node, font=\large]
     at (146.649, 636) {$r$};
  \draw
    (148, 644)
     -- (148, 660)
     -- (244, 660)
     -- (244, 660);
  \node[ipe node, font=\large]
     at (161.568, 641.351) {$\mathcal{F}_{\theta,\phi}$};
\end{tikzpicture}

%% file: Feedforward_parametrization2.tex
\tikzstyle{ipe stylesheet} = [
  ipe import,
  even odd rule,
  line join=round,
  line cap=butt,
  ipe pen normal/.style={line width=0.4},
  ipe pen heavier/.style={line width=0.8},
  ipe pen fat/.style={line width=1.2},
  ipe pen ultrafat/.style={line width=2},
  ipe pen normal,
  ipe mark normal/.style={ipe mark scale=3},
  ipe mark large/.style={ipe mark scale=5},
  ipe mark small/.style={ipe mark scale=2},
  ipe mark tiny/.style={ipe mark scale=1.1},
  ipe mark normal,
  /pgf/arrow keys/.cd,
  ipe arrow normal/.style={scale=7},
  ipe arrow large/.style={scale=10},
  ipe arrow small/.style={scale=5},
  ipe arrow tiny/.style={scale=3},
  ipe arrow normal,
  /tikz/.cd,
  ipe arrows, 
  <->/.tip = ipe normal,
  ipe dash normal/.style={dash pattern=},
  ipe dash dotted/.style={dash pattern=on 1bp off 3bp},
  ipe dash dashed/.style={dash pattern=on 4bp off 4bp},
  ipe dash dash dotted/.style={dash pattern=on 4bp off 2bp on 1bp off 2bp},
  ipe dash dash dot dotted/.style={dash pattern=on 4bp off 2bp on 1bp off 2bp on 1bp off 2bp},
  ipe dash normal,
  ipe node/.append style={font=\normalsize},
  ipe stretch normal/.style={ipe node stretch=1},
  ipe stretch normal,
  ipe opacity 10/.style={opacity=0.1},
  ipe opacity 30/.style={opacity=0.3},
  ipe opacity 50/.style={opacity=0.5},
  ipe opacity 75/.style={opacity=0.75},
  ipe opacity opaque/.style={opacity=1},
  ipe opacity opaque,
]
\definecolor{red}{rgb}{1,0,0}
\definecolor{blue}{rgb}{0,0,1}
\definecolor{green}{rgb}{0,1,0}
\definecolor{yellow}{rgb}{1,1,0}
\definecolor{orange}{rgb}{1,0.647,0}
\definecolor{gold}{rgb}{1,0.843,0}
\definecolor{purple}{rgb}{0.627,0.125,0.941}
\definecolor{gray}{rgb}{0.745,0.745,0.745}
\definecolor{brown}{rgb}{0.647,0.165,0.165}
\definecolor{navy}{rgb}{0,0,0.502}
\definecolor{pink}{rgb}{1,0.753,0.796}
\definecolor{seagreen}{rgb}{0.18,0.545,0.341}
\definecolor{turquoise}{rgb}{0.251,0.878,0.816}
\definecolor{violet}{rgb}{0.933,0.51,0.933}
\definecolor{darkblue}{rgb}{0,0,0.545}
\definecolor{darkcyan}{rgb}{0,0.545,0.545}
\definecolor{darkgray}{rgb}{0.663,0.663,0.663}
\definecolor{darkgreen}{rgb}{0,0.392,0}
\definecolor{darkmagenta}{rgb}{0.545,0,0.545}
\definecolor{darkorange}{rgb}{1,0.549,0}
\definecolor{darkred}{rgb}{0.545,0,0}
\definecolor{lightblue}{rgb}{0.678,0.847,0.902}
\definecolor{lightcyan}{rgb}{0.878,1,1}
\definecolor{lightgray}{rgb}{0.827,0.827,0.827}
\definecolor{lightgreen}{rgb}{0.565,0.933,0.565}
\definecolor{lightyellow}{rgb}{1,1,0.878}
\definecolor{black}{rgb}{0,0,0}
\definecolor{white}{rgb}{1,1,1}
\begin{tikzpicture}[ipe stylesheet]
  \filldraw[ipe dash dashed, fill=gray, ipe opacity 30]
    (80, 820) rectangle (268, 732);
  \filldraw[black, ipe opacity 10]
    (88.4123, 793.506)
     -- (88.3743, 733.77)
     -- (246.417, 733.333)
     -- (246.541, 793.721)
     -- cycle;
  \filldraw[black, ipe opacity 10]
    (87.994, 817.986)
     -- (88.0555, 798.074)
     -- (246.073, 797.907)
     -- (246.018, 818.146)
     -- cycle;
  \draw
    (116, 816) rectangle (196, 800);
  \node[ipe node]
     at (120, 804) {$\sum_{i=1}^{N_a} a_i g_y(\psi_i(r))$};
  \draw
    (212, 816) rectangle (244, 800);
  \node[ipe node]
     at (214.128, 806.08) {$\frac{1}{B(q^{-1})}$};
  \draw[->]
    (196, 808)
     -- (212, 808);
  \draw
    (100.022, 790.012)
     -- (100.022, 778.059)
     -- (120.1897, 777.992)
     -- (120.1897, 789.992)
     -- cycle;
  \node[ipe node]
     at (107.221, 781.157) {$1$};
  \draw
    (99.9271, 769.996)
     -- (99.8323, 758.043)
     -- (120.1897, 757.992)
     -- (120.1897, 769.992)
     -- cycle;
  \node[ipe node]
     at (102.462, 760.967) {$q^{-1}$};
  \draw[->]
    (92, 764)
     -- (92, 784)
     -- (100, 784);
  \draw[->]
    (92, 764)
     -- (92, 744)
     -- (100, 744);
  \draw
    (136, 784.208) circle[radius=8];
  \node[ipe node]
     at (130.879, 781.438) {$h_1^0$};
  \draw
    (136, 764) circle[radius=8];
  \node[ipe node]
     at (130.879, 761.438) {$h_2^0$};
  \draw
    (136, 744) circle[radius=8];
  \node[ipe node]
     at (130.879, 741.438) {$h_3^0$};
  \draw
    (168, 784) circle[radius=8];
  \node[ipe node]
     at (190.674, 761.138) {$h_1^2$};
  \draw
    (168, 764) circle[radius=8];
  \node[ipe node]
     at (162.879, 781.438) {$h_1^1$};
  \draw
    (168, 744) circle[radius=8];
  \node[ipe node]
     at (162.879, 741.438) {$h_3^1$};
  \draw
    (196, 764) circle[radius=8];
  \draw[->]
    (144, 744)
     -- (160, 784);
  \draw[->]
    (144, 744)
     -- (160, 764);
  \draw[->]
    (144, 744)
     -- (160, 744);
  \draw[->]
    (144, 784)
     -- (160, 784);
  \draw[->]
    (144, 784)
     -- (160, 764);
  \draw[->]
    (144, 784)
     -- (160, 744);
  \draw[->]
    (144, 764)
     -- (160, 764);
  \draw[->]
    (144, 764)
     -- (160, 784);
  \draw[->]
    (144, 764)
     -- (160, 744);
  \draw
    (99.9271, 749.996)
     -- (99.8323, 738.043)
     -- (120.1897, 737.992)
     -- (120.1897, 749.992)
     -- cycle;
  \node[ipe node]
     at (102.462, 740.967) {$q^{-2}$};
  \draw[->]
    (84, 784)
     -- (84, 764)
     -- (100, 764);
  \draw
    (120, 784)
     -- (128, 784);
  \draw
    (120, 764)
     -- (128, 764);
  \draw
    (120, 744)
     -- (128, 744);
  \draw[->]
    (176, 764)
     -- (188, 764);
  \draw[->]
    (176, 784)
     -- (188, 764);
  \draw[->]
    (176.0949, 743.9051)
     -- (188.0949, 763.9051);
  \draw
    (212, 772) rectangle (244, 756);
  \node[ipe node]
     at (214.128, 762.08) {$\frac{1}{B(q^{-1})}$};
  \draw[->]
    (204, 764)
     -- (212, 764);
  \node[ipe node]
     at (233.001, 785.231) {$\mathcal{C}_\phi$};
  \node[ipe node]
     at (71.367, 787.018) {$r$};
  \draw[->]
    (72, 784)
     -- (84, 784)
     -- (84, 808)
     -- (116, 808);
  \draw
    (252, 784) circle[radius=4];
  \draw
    (248, 784)
     -- (256, 784);
  \draw
    (252, 780)
     -- (252, 788);
  \draw[->]
    (244, 808)
     -- (252, 808)
     -- (252, 788);
  \draw[->]
    (256, 784)
     -- (276, 784);
  \draw[->]
    (244, 764)
     -- (252, 764)
     -- (252, 780);
  \node[ipe node]
     at (248, 753.921) {$f_\mathcal{C}$};
  \node[ipe node]
     at (248, 812) {$f_\mathcal{M}$};
  \node[ipe node]
     at (271.947, 772.996) {$f$};
  \node[ipe node]
     at (247.584, 737.088) {$\mathcal{F}_{\theta,\phi}$};
  \node[ipe node]
     at (89.255, 810.087) {$\mathcal{M}_\theta$};
  \draw
    (161.8574, 761.7994)
     .. controls (164.7744, 761.6424) and (165.9254, 762.0714) .. (166.6927, 762.5723)
     .. controls (167.4601, 763.0731) and (167.8437, 763.6458) .. (168.2272, 764.2188)
     .. controls (168.6107, 764.7918) and (168.9941, 765.3651) .. (169.7612, 765.8663)
     .. controls (170.5284, 766.3674) and (171.6794, 766.7964) .. (174.2224, 766.7454);
\end{tikzpicture}

%% file: Sections/SK_iterations.tex
\section{SK Iterations for Output Error Minimization }
\label{sec:SK_iterations}
In this section, an output error criterion is introduced to be minimized by the learned parameters $\theta,\phi$ of $\mathcal{F}_{\theta,\phi}$ in \eqref{eq:ffw_parametrization} (contribution C2). This criterion can be seen as a sequence of weighted least-squares problems, known a SK-iterations. 

The OE criterion directly penalizes deviations of $f(k)$ from $\hat{f}(k)$ to ensure that $f(k) = \hat{f}(k)$\textcolor{red}{\ifthenelse{\boolean{shorten}}{\sout{ for perfect feedforward,}}{}}, as defined next.
\begin{definition}
	Given feedforward parametrization \eqref{eq:ffw_parametrization} and dataset $\mathcal{D}$, the OE criterion $J_{OE} \in \mathbb{R}_{\geq 0}$ is given by
	\begin{align}
		J_{OE} &= \sum_{k=1}^N \left( \hat{f}(k) - \frac{1}{B(q)}  \left( A(r(k)) + g_\phi(r(k)) \right) \right)^2\hspace{-5pt}, \label{eq:J_OE}
	\end{align}
	in which $(B(q))^{-1}(\cdot)$ represents a filtering operation.
	\label{def:J_OE}
\end{definition}
Criterion \eqref{eq:J_OE} is linear in the parameters $a$ of the exogenous dynamics, but nonlinear in the parameters $b$ of the AR dynamics. 
As a result, \eqref{eq:J_OE} is nonconvex in $b$. 

This nonconvexity in $b$ can also be regarded as an a priori unknown weighting function of a least-squares problem. More specifically, \eqref{eq:J_OE} can be written as
\begin{equation}
	\resizebox{.91\hsize}{!}{$J_{OE} = \sum\limits_{k=1}^N \left( \frac{1}{B(q)} ( B(q) \hat{f}(k) - A(r(k)) - g_\phi(r(k)) ) \right)^2 \hspace{-3pt}.$}
	\label{eq:J_OE_2}
\end{equation}
Criterion $J_{OE}$ in \eqref{eq:J_OE_2} is still nonlinear in parameters $b$ due to the filtering term $(B(q))^{-1}$, but is linear in $b$ in the term $B(q) \hat{f}(k)$. Thus, given the weighting function $(B(q))^{-1}$, the problem is linear in $\theta = \textrm{col}(a,b)$. This motivates the following optimization algorithm for $J_{OE}$.

\vspace{3pt}
\hrule
\vspace{-2pt}
\begin{procedure}[SK-iterations for OE optimization] \hfill
	\vspace{3pt} \hrule \vspace{2pt}
	Given parametrization \eqref{eq:ffw_parametrization} with parameters $a$, $b$, $\phi$, and dataset $\mathcal{D}$, set $j=1$ and initialize $a^{0},b^{0},\phi^{0}$ according to some strategy (e.g., $a^{0},b^{0}$ as the best linear approximation, and $\phi^{0}$ through Glorot initialization \cite{Glorot2010}). Then,\textcolor{red}{\ifthenelse{\boolean{shorten}}{\sout{ carry out the following steps:}}{}} iterate:
	\begin{enumerate}[label=(\arabic*)]
		\item Given $B^{j-1}(q)$, determine $a^{j}, b^{j}, \phi^{j}$ as
		\begin{equation}
		\begin{aligned}
			a^{j}, b^{j}, \phi^{j} = \arg \min_{a,b,\phi} J_{OE}^j,
		\end{aligned}
		\end{equation}
		with $J_{OE}^j \in \mathbb{R}_{\geq 0}$ given by
		\begin{equation}
		\begin{aligned}
		J_{OE}^j = \sum_{k=1}^N \left( \frac{1}{B^{j-1}(q)} \left( B(q) \hat{f}(k) \right. \right. \\ \left. \left. - A(r(k)) - g_\phi(r(k)) \right) \right)^2.
		\end{aligned} \label{eq:J_OE_SK}
		\end{equation}	
		\item Set $j = j + 1$ and go back to (1) until convergence, e.g., until $a^{j} = a^{j-1}$, $b^{j} = b^{j-1}$, $\phi^{j} =\phi^{j-1}$.
	\end{enumerate}
	\label{proc:SK_iterations}
	\vspace{0pt} 	\hrule 	\vspace{-2pt}
\end{procedure}
The minimization \eqref{eq:J_OE_SK} can be carried out through standard optimizers\textcolor{red}{\ifthenelse{\boolean{shorten}}{\sout{ such as ADAM}}{}} by differentiating through\textcolor{red}{\ifthenelse{\boolean{shorten}}{\sout{ the filtering operation}}{}} $(B^{j-1}(q))^{-1}$.

In \eqref{eq:J_OE_SK} and Algorithm \ref{proc:SK_iterations}, $(B(q))^{-1}$ is interpreted as an a priori unknown weighting function that is iteratively adjusted over the iterations. Through iterating over $j$, it is aimed to recover \eqref{eq:J_OE_2} when $B^{j-1}(q) = B^{j}(q)$. Despite the lack of theoretical convergence guarantees and the nonconvexity of \eqref{eq:J_OE}, practical use of this SK algorithm has shown good convergence properties \cite{Whitfield1987, 6837472}.


%% file: Sections/non_uniqueness_EE_LS.tex
\section{Orthogonal Projection-Based Regularizer}
\label{sec:OP_regularizer}
Since all iterations of Algorithm \ref{proc:SK_iterations} for optimizing $J_{OE}$ in \eqref{eq:J_OE} are the same up to the weighting $(B^{j-1}(q))^{-1}$, the first iteration $J_{OE}^1$ is analyzed for the simplified setting in which only the last layer of $g_\phi$ in \eqref{eq:FNN} is optimized. The optimum corresponding to this simplified problem is often non-unique due to the universal approximator characteristics of $g_\phi$. In this paper, an orthogonal projection-based regularization is used to ensure that the optimum for the model coefficients $\theta$ is unique (contribution C3). This non-uniqueness directly applies to the full case \eqref{eq:J_OE}.
\subsection{Non-Uniqueness of First SK Iteration}
If only the last layer of $g_\phi$ in \eqref{eq:FNN} is optimized, $g_\phi$ is also LIP, such that the first SK-iteration can be written as a convex least-squares problem. The solution to this least-squares problem is often non-unique due to the universal approximation characteristics of $g_\phi$. More specifically, consider criterion $J_{OE}^1$ in \eqref{eq:J_OE_SK} with $B^{0}(q) = 1$ defined below.
	\begin{definition}
	Given feedforward parametrization \eqref{eq:ffw_parametrization} and dataset $\mathcal{D}$, $J_{OE}^1$ with $B^{0}(q) = 1$ is given by
	\begin{equation}
		J_{OE}^1 = \sum_{k=1}^N \left( B(q)\hat{f}(k) - A(r(k)) - g_\phi(r(k)) \right)^2 \label{eq:J_EE}.
	\end{equation}
\end{definition}
\begin{remark}
	This criterion can be recognized as the equation error corresponding to feedforward parametrization \eqref{eq:ffw_parametrization}.
\end{remark}
Consider now the case in which all hidden layers of $g_\phi$ in \eqref{eq:FNN} are fixed, and only the output layer is optimized, i.e.,
\begin{align}
	g_\phi(r(k)) = h^L(r(k))^\T \phi^\T, \label{eq:LIP_approximator}
\end{align}
with $\phi = W^L \in \mathbb{R}^{1 \times N_\phi}$. For this setting, the approximator is also LIP, which allows to rewrite criterion \eqref{eq:J_EE} as follows.
\begin{lemma}
	Given an approximator structure \eqref{eq:LIP_approximator}, $J_{OE}^1$ in \eqref{eq:J_EE} can be represented as
	\begin{align}
		J_{OE}^1 = \norm{\fhl - M \theta - H(\rl)^\T \phi^\T}{2}^2, \label{eq:J_EE_lifted}
	\end{align}
	where $\theta = \begin{bmatrix}a^\T & b^\T \end{bmatrix}$ and $M = \begin{bmatrix} R & -\hat{F} \end{bmatrix}$ with
	\begin{align}
		\fhl &= \begin{bmatrix} \hat{f}(1) & \hat{f}(2) & \ldots & \hat{f}(N) \end{bmatrix}^\T \in \mathbb{R}^N \nonumber \\
		\hat{F} &= \begin{bmatrix} q^{-1} \fhl & q^{-2} \fhl & \ldots & q^{-N_b} \fhl \end{bmatrix} \in \mathbb{R}^{N \times N_b} \nonumber \\
	R &= \begin{bmatrix} g_1(\psi_1(\rl)) & \ldots & g_{N_a}(\psi_{N_a}(\rl)) \end{bmatrix} \in \mathbb{R}^{N \times N_a} \label{eq:J_EE_lifted_definitions} \\
	H(\rl) &= \begin{bmatrix} h^{L}(r(1)) & \ldots & h^{L}(r(N)) \end{bmatrix} \in \mathbb{R}^{N_\phi \times N} \nonumber,
	\end{align}
	in which $\psi_i(\rl) = \begin{bmatrix}	(\psi_i(r))(1) & \hdots & (\psi_i(r))(N) \end{bmatrix}^\T \in \mathbb{R}^{N}$ and $g_i$ applies elementwise.
	\label{lem:J_EE_lifted}
\end{lemma}
Criterion \eqref{eq:J_EE_lifted} is a standard least-squares problem for which the solution is given by the pseudoinverse.
\begin{lemma}
	\label{lem:pinv_solution}
Given $\begin{bmatrix}M & H(\rl)^\T \end{bmatrix} \in \mathbb{R}^{N \times N_\theta + N_\phi}$, the minimizer $\theta^*,\phi^*$ of $J_{OE}^1$ in \eqref{eq:J_EE_lifted} is given by
\begin{equation}
	\theta^*,\phi^* = \arg \min_{\theta,\phi} J_{OE}^1 = \begin{bmatrix}M & H(\rl)^\T \end{bmatrix}^+ \fhl + \begin{bmatrix} v_\theta \\ v_\phi \end{bmatrix},
	\label{eq:J_EE_minimizer}
\end{equation}
for any $v = \begin{bmatrix}v_\theta^\T & v_\phi^\T \end{bmatrix}^\T \in \mathbb{R}^{N_\theta + N_\phi}$ such that $v \in \textnormal{ker} \begin{bmatrix}M & H(\rl)^\T \end{bmatrix}$, where $(\cdot)^+$ represents the pseudoinverse.
\end{lemma}
Even though $\begin{bmatrix}M & H(\rl)^\T \end{bmatrix}$ is tall, i.e., $N > N_\theta + N_\phi$, $\textnormal{ker} \begin{bmatrix}M & H(\rl)^\T \end{bmatrix}$ can be non-empty by two mechanisms. Before discussing these mechanisms, the following is assumed.
\begin{assumption}
	\textcolor{red}{\ifthenelse{\boolean{shorten}}{\sout{The matrix representation $M \in \mathbb{R}^{N \times N_\theta}$ in \eqref{eq:J_EE_lifted_definitions} has full rank.}}{}} For tall $M \in \mathbb{R}^{N \times N_\theta}$, $\textrm{rank}\ M = N_{\theta}$.
	\label{ass:M_full_rank}
\end{assumption}
This assumption corresponds to a persistence of excitation condition for the model parametrization \eqref{eq:model_class}. For $g_i(\cdot) = \textrm{Id}(\cdot)$, i.e., for rational model parametrizations, this is equivalent to a non-zero spectrum of $r$ at $N_\theta$ points \cite{sysID_Ljung}. Assumption \ref{ass:M_full_rank} now allows for the following lemma.

\begin{lemma}
	$\textnormal{ker} \begin{bmatrix}M & H(\rl)^\T \end{bmatrix}$ is nonempty if and only if one of the following conditions is satisfied.
	\begin{enumerate}[label=P\arabic*)]
	\item There exists $v_\phi$ for which $H(\rl)^\T v_\phi = 0$, and $\begin{bmatrix} 0 & v_\phi^\T \end{bmatrix}^\T \in \textnormal{ker} \begin{bmatrix}M & H(\rl)^\T \end{bmatrix}$. 
	\item There exists a column $M_i \in \im\ H(\rl)^\T $. Consequently, there exists a $v$ such that $\begin{bmatrix} M & H(\rl)^\T \end{bmatrix} v = 0$.
\end{enumerate} 
\label{lem:modes_of_non_uniqueness}
\end{lemma}
The case $P1$ corresponds to overparametrization of $g_\phi$, and only results in non-unique approximator coefficients $\phi$, which do not need to be interpretable, and is thus of no concern. In the case of $P2$, $g_\phi$ can represent (parts of) the model due to its universal function approximator characteristics\textcolor{red}{\ifthenelse{\boolean{shorten}}{\sout{. Numerical examples show that this mechanism}}{}}which can be present in practice\cite{Kon2022PhysicsGuided}. In this case, the model coefficients $\theta$ are not unique.
\subsection{Orthogonal Decomposition}
An explicit expression describing the subspace in which $\theta$ is non-unique is obtained through splitting the criterion \eqref{eq:J_EE_lifted} into orthogonal subspaces, which are chosen as the model output space $\im\ M$, and its orthogonal complement. 

\textcolor{red}{\ifthenelse{\boolean{shorten}}{\sout{More specifically, consider criterion \eqref{eq:J_EE_lifted}. Given that $M$ has full rank, see Assumption \ref{ass:M_full_rank}, it can be factorized through a singular value decomposition (SVD) as follows.}}{}}More specifically, given that $M$ has full rank, it can be factorized through a singular value decomposition (SVD).
\begin{lemma}
	\textcolor{red}{\ifthenelse{\boolean{shorten}}{\sout{The full-rank matrix }}{}}$M \in \mathbb{R}^{N \times N_\theta}$, $N > N_\theta$, can be factorized as
	\begin{equation}
		M = \begin{bmatrix} U_1 & U_2 \end{bmatrix} \begin{bmatrix} \Sigma \\ 0	\end{bmatrix} V^\T,
		\label{eq:SVD}
	\end{equation}
	with $U_1 \in \mathbb{R}^{N \times N_\theta}$, $U_2 \in \mathbb{R}^{N \times N - N_\theta}$, $V \in \mathbb{R}^{N_\theta \times N_\theta}$ unitary matrices such that
	$U_1^\T U_1 = I_{N_\theta}$, $U_1^\T U_2 = 0$, $U_1 U_1^\T + U_2 U_2^\T = I_N$, and $\Sigma \in \mathbb{R}^{N_\theta \times N_\theta} = \textrm{diag}(\sigma_1, \ldots, \sigma_{N_\theta})$ with $\sigma_i > 0$ \cite{lay2003linear}.
	\label{lem:SVD}
\end{lemma}
Consequently, the model response $M \theta$ can be written as
\begin{equation}
	M \theta = U_1 \Sigma V^\T \theta,
	\label{eq:SVD_model_resp}
\end{equation}
in which $U_1$ is a basis for the output space of $M$, and $U_2$ its orthogonal complement. This explicit basis allows to decouple criterion \eqref{eq:J_EE_lifted} into orthogonal subspaces\textcolor{red}{\ifthenelse{\boolean{shorten}}{\sout{, as formalized next}}{}}.
\begin{theorem}
	\label{th:J_EE_lifted_decoupled}
	Given factorization \eqref{eq:SVD}, $J_{OE}^1$ in \eqref{eq:J_EE_lifted} can be written as  
	\begin{equation}
	J_{OE}^1 = \left\lVert
	\begin{bmatrix}
		U_1^\T \fhl 
		\\ U_2^\T \fhl 
	\end{bmatrix}
	-
	\begin{bmatrix}
		\Sigma V^\T & U_1^\T H(\rl)^\T \\
		0 & U_2^\T H(\rl)^\T
	\end{bmatrix}
	\begin{bmatrix}
		\theta \\
		\phi^\T
	\end{bmatrix}
	\right\rVert_2^2 \label{eq:J_EE_lifted_decoupled}.
	\end{equation}%
\end{theorem}%
This decoupling can be interpreted as projection into the model coefficient space and into its orthogonal complement. The entry $U_1^\T H(\rl)^\T \phi^\T$ represents the contribution of the approximator expressed in the coordinates of model coefficients. Theorem \ref{th:J_EE_lifted_decoupled} allows for the following result.
\begin{corollary}
	\label{cor:singular_subspace}
	Given\textcolor{red}{\ifthenelse{\boolean{shorten}}{ \sout{the optimizer to $J_{OE}^1$ in \eqref{eq:J_EE_minimizer}}}{} }\eqref{eq:J_EE_minimizer}, if a vector $v = \begin{bmatrix}v_\theta^\T & v_\phi^\T \end{bmatrix}^\T$ exists such that $\begin{bmatrix} M & H(\rl)^\T \end{bmatrix} v = 0$, then $v$ satisfies
	\begin{equation}
	\begin{bmatrix}
		\Sigma V^\T & U_1^\T H(\rl)^\T \\
		0 & U_2^\T H(\rl)^\T
	\end{bmatrix}
	\begin{bmatrix}
		v_\theta \\
		v_\phi
	\end{bmatrix} 
	= \begin{bmatrix} 0 \\ 0	\end{bmatrix},
	\label{eq:matrix_subspace}
	\end{equation}
	such that $H(\rl)^\T v_\phi \in \textnormal{ker}\ U_2^T = (\im\ U_2)^\perp = \im\ U_1$, and
	\begin{equation}
		v_\theta = -(\Sigma V^\T )^{-1} U_1^\T H(\rl)^\T v_\phi = -M^+ H(\rl)^T v_\phi.	\label{eq:singular_subspace_expression}
	\end{equation}
\end{corollary}
The case where $H(\rl)^\T v_\phi = 0$ for $v_\phi \neq 0$ corresponds to $P1$ of Lemma \ref{lem:modes_of_non_uniqueness}. In contrast, $H(\rl)^\T v_\phi \neq 0$ and $H(\rl)^\T v_\phi \in \im\ U_1$ corresponds to $P2$, i.e., there exists a linear subspace in which both the model $M$ and approximator $H(\rl)^\T$ can capture the same effects. Corollary \ref{cor:singular_subspace} expresses the relation between $v_\theta$ and $v_\phi$ for any $v$ in this subspace, describing the directions in which $\theta$ is non-unique.

\subsection{Orthogonal Projection-Based Regularizer}
To obtain unique model coefficients $\theta$, $J_{OE}^1$ in \eqref{eq:J_EE_lifted} is regularized with an orthogonal projection-based regularization that penalizes the approximator output $H(\rl)^\T \phi^\T$ in the subspace of the model $M \theta$. This orthogonal projection-based cost function for $J_{OE}^1$ where $g_\phi$ is LIP is defined next.

\begin{definition}
	Given dataset $\mathcal{D}$ and $J_{OE}^1$ in \eqref{eq:J_EE_lifted}, the criterion $J_{OE,P}^1 \in \mathbb{R}_{\geq 0}$ is defined as
	\begin{equation}
		J_{OE,P}^1 = \norm{\fhl - M \theta - H(\rl)^\T \phi^\T}{2}^2 + \lambda R(\phi),
		\label{eq:J_EE_lifted_OP_reguralized}
	\end{equation}
	in which $R(\phi) \in \mathbb{R}_{\geq 0}$ is given by
	\begin{equation}
		R(\phi) = \norm{(\Sigma V^\T)^{-1} U_1^\T H(\rl)^\T \phi^\T}{2}^2.
		\label{eq:orthogonal_regularizer_EE}
	\end{equation}
\end{definition}
The regularizer $R(\phi)$ penalizes the \textit{scaled} approximator output $H(\rl)^\T \phi^\T$ in $\im\ M= \im\ U_1$ through $U_1^T H(\rl)^\T \phi^\T$. Through the scaling $(\Sigma V^\T)^{-1}$, $R(\phi)$ directly regularizes for $v_\theta = 0$, see \eqref{eq:singular_subspace_expression}.
The structure of \eqref{eq:orthogonal_regularizer_EE} allows for splitting \eqref{eq:J_EE_lifted_OP_reguralized} into orthogonal subspaces as formalized next.
\begin{theorem}
	\label{th:J_EE_lifted_OP_decouped}
	Given factorization \eqref{eq:SVD}, $J_{OE,P}^1$ in \eqref{eq:J_EE_lifted_OP_reguralized} can be written as
	{\small
	\begin{equation}
	J_{OE,P}^1 = \left\lVert \hspace{-1pt}
	\begin{bmatrix}
		U_1^\T \fhl \\ 
		U_2^\T \fhl \\
		0
	\end{bmatrix}
	-
	\begin{bmatrix}
		\Sigma V^\T & U_1^\T H(\rl)^\T \\
		0 & U_2^\T H(\rl)^\T \\
		0 & \sqrt{\lambda}(\Sigma V^\T)^{-1} U_1^\T H(\rl)^\T
	\end{bmatrix}
	\hspace{-1pt}
	\begin{bmatrix}
		\theta \\
		\phi^\T
	\end{bmatrix}
	\hspace{-1pt} \right\rVert_2^2 \hspace{-4pt}.
	\label{eq:J_EE_lifted_OP_decoupled}
	\end{equation}}
\end{theorem}
Theorem \ref{th:J_EE_lifted_OP_decouped} shows that the regularizer \eqref{eq:orthogonal_regularizer_EE} adds additional rows to the decoupled optimization compared to \eqref{eq:J_EE_lifted_decoupled} of Theorem \ref{th:J_EE_lifted_decoupled}. These extra rows ensure that unique model coefficients $\theta$ are recovered from $J_{OE,P}^1$, as illustrated next.
\begin{corollary}
\label{cor:unique_theta}
	Given criterion $J_{OE}^1$ in \eqref{eq:J_EE_lifted_decoupled} and $J_{OE,P}^1$ in \eqref{eq:J_EE_lifted_OP_decoupled}, nominal solution $\begin{bmatrix}M & H(\rl)^\T \end{bmatrix}^+ \fhl := x^*$, see Lemma \ref{lem:pinv_solution}, and any two vectors $v_1, v_2 \in \textnormal{ker} \begin{bmatrix}M & H(\rl)^\T \end{bmatrix}$ such that $v_1 = \begin{bmatrix} 0 & v_\phi^\T \end{bmatrix}^\T$ and $v_2 = \begin{bmatrix} v_\theta^\T & v_\phi^\T \end{bmatrix}^\T$ with $v_\theta \neq 0$, then,
	\begin{equation}
		J_{OE}^1(x^* + v_1) = J_{OE}^1(x^* + v_2).
	\end{equation}
	In contrast, for $J_{OE,P}^1$, it holds that
	\begin{equation}
		J_{OE,P}^1(x^* + v_1) < J_{OE,P}^1(x^* + v_2),
	\end{equation}
	such that $\theta^*$ in $\arg \min_{\theta,\phi} J_{OE,P}^1$ is unique.
\end{corollary}
Corollary \ref{cor:unique_theta} conveys that the orthogonal projection-based regularizer \eqref{eq:orthogonal_regularizer_EE} shrinks the non-unique directions $v_\theta$ to the zero vector: for any vector $\begin{bmatrix} v_\theta^\T & v_\phi^\T \end{bmatrix}^\T \in \textnormal{ker } \begin{bmatrix} M & H(\rl)^\T \end{bmatrix}$, the $v_\theta$ component is regularized to 0, such that unique model coefficients $\theta$ are recovered. Note that the contribution $v_\phi$ can still be non-unique, i.e., $P1$ of Lemma \ref{lem:modes_of_non_uniqueness}.


\begin{remark}
	Other regularization techniques,\textcolor{red}{\ifthenelse{\boolean{shorten}}{\sout{ such as $\ell_2$, $\ell_1$ or Tikinov regularization}}{}}e.g., $\ell_2$, could have also been employed to obtain unique $\theta$ in \eqref{eq:J_EE_lifted_decoupled}. However, $R(\phi)$ in \eqref{eq:orthogonal_regularizer_EE} \textcolor{red}{\ifthenelse{\boolean{shorten}}{\sout{promotes unique $\theta$ through penalizing outputs of $g_\phi$ that can be captured by $M\theta$, thus simply shifting contributions, whereas other regularizations also introduce coefficient bias.}}{}}only penalizes outputs of $g_\phi$ that can be captured by $M\theta$, whereas others also penalize outputs that can only be captured by $g_\phi$, resulting in performance decrease.
\end{remark}

This section has shown that the optimum of $J_{OE}^1$ in \eqref{eq:J_EE} is non-unique already when only the last layer of $g_\phi$ in \eqref{eq:FNN} is optimized. 
Naturally, this problem persists if all layers of $g_\phi$ are optimized, for which the linear subspace $\eqref{eq:singular_subspace_expression}$ becomes a complex nonlinear manifold in $\mathbb{R}^{N_\theta + N_\phi}$. Also in this full setting, $R(\phi)$ promotes unique $\theta$ for $J_{OE}^1$. This regularization is extended to subsequent SK-iterations in the next section.



%% file: Sections/SK_and_Orthogonal_regularization.tex
\section{Orthogonality at Each SK-Iteration}
\label{sec:SK_and_orthogonal_reg}
In this section, the orthogonal projection-based regularizer \eqref{eq:orthogonal_regularizer_EE},\textcolor{red}{\ifthenelse{\boolean{shorten}}{ \sout{see Section \ref{sec:OP_regularizer}},}{}} is incorporated in the SK-iterations of Algorithm \ref{proc:SK_iterations}, see Section \ref{sec:SK_iterations}, resulting in an efficient solver for OE minimization that promotes uniqueness of $\theta$ at each iteration.

This uniqueness is achieved through an iteration-varying orthogonal projection-based regularizer. This regularizer is obtained through constructing an orthogonal decomposition of the weighted model response alike to Lemma \ref{lem:SVD}. Then, $J_{OE}^j$ is regularized similarly to \eqref{eq:J_EE_lifted_OP_reguralized}, such that it can be decoupled like \eqref{eq:J_EE_lifted_OP_decoupled} at each iteration. Here, due to space constraints, only the resulting algorithm is presented.

\vspace{3pt}
\hrule
\vspace{-2pt}
\begin{procedure}[SK-iterations for OE minimization with orthogonal projection-based regularization] \hfill
	\vspace{3pt} \hrule \vspace{2pt}
	Given parametrization \eqref{eq:ffw_parametrization} with parameters $a$, $b$, $\phi$, and dataset $\mathcal{D}$, set $j=1$ and initialize $a^{0},b^{0},\phi^{0}$. Then,\textcolor{red}{\ifthenelse{\boolean{shorten}}{\sout{carry out the following steps}}{} }iterate:
	\begin{enumerate}[label=(\arabic*)]
		\item Given $B^{j-1}(q)$, calculate its convolution matrix $W^{j-1}$ such that the finite-time response $y(k) = (B^{j-1}(q))^{-1} u(k)$ is given by $\underline{y} = W^{j-1} \underline{u}$ with
	{\small
	\begin{equation}
		W^{j-1} = \begin{bmatrix}	
		h(0) & h(-1) & \ldots & h(1-N) \\
		h(1) & h(0) & \ldots & h(2-N) \\
		\vdots & & \ddots & \vdots \\
		h(N-1) & h(N-2) & \ldots & h(0)
	\end{bmatrix},
	\label{eq:conv_mat}
	\end{equation}
	}%
	with\textcolor{red}{\ifthenelse{\boolean{shorten}}{\sout{ $h(k), k \in \mathbb{Z}$}}{}} $h(k)$ the\textcolor{red}{\ifthenelse{\boolean{shorten}}{\sout{impulse response coefficients}}{} }impulse response of $(B^{j-1}(q))^{-1}$.
		\item Rewrite $J_{OE}^j$ in \eqref{eq:J_OE_SK} as a vector norm, i.e.,
		\begin{equation}
			J_{OE}^j = \norm{ W^{j-1}  \left(\fhl - M \theta - g_\phi(\rl) \right)}{2}^2. \label{eq:J_OE_SK_lifted}
		\end{equation}
		\item Obtain the SVD of $W^{j-1} M$ as
		\begin{equation}
		W^{j-1} M = \begin{bmatrix} U_1^{j-1} & U_2^{j-1} \end{bmatrix} \begin{bmatrix} \Sigma^{j-1} \\ 0	\end{bmatrix} V^{j-1^\T}.
	\label{eq:SVD_W_M}
	\end{equation}
		\item Construct iteration-varying orthogonal projection-based regularizer $R^{j-1}(\phi)$ as 
		\begin{equation}
		\hspace{0pt}\resizebox{.85\hsize}{!}{$R^{j-1}(\phi) = \norm{(\Sigma^{j-1} V^{j-1^\T})^{-1} U_1^{j-1^\T} W^{j-1} g_\phi(\rl)}{2}^2.$}
		\label{eq:orthogonal_regularizer_OE_SK}
	\end{equation}
		\item Determine $a^{j}, b^{j}, \phi^{j}$ as	
		\begin{equation}
			a^{j}, b^{j}, \phi^{j} \arg \min_{a,b,\phi} = J_{OE}^j + \lambda R^{j-1}(\phi).
		\end{equation}
		\item Set $j = j + 1$ and go back to (1) until convergence\textcolor{red}{\ifthenelse{\boolean{shorten}}{\sout{, e.g., $a^{j} = a^{j-1}$, $b^{j} = b^{j-1}$, $\phi^{j} =\phi^{j-1}$}}{}}.
	\end{enumerate}
	\label{proc:SK_iterations_with_OP_regularizer}
	\vspace{-1pt} 	\hrule 	\vspace{-4pt}
\end{procedure}
In this regularized SK-algorithm, $R^{j-1}(\phi)$ directly promotes uniqueness of $\theta$ at each iteration through optimizing $g_\phi$ such that $W^{j-1} g_\phi(\rl) \notin \im  \ U_1^{j-1}$, and consequently $W^{j-1} M \theta$ captures all effects that can be encapsulated by the model. Thus, heuristically, $\theta$ is unique at convergence, resulting in unique model coefficients for\textcolor{red}{\ifthenelse{\boolean{shorten}}{\sout{ feedforward parametrization}}{}} \eqref{eq:ffw_parametrization}.

%% file: Sections/Simulation_validation.tex
\section{Simulation Example}
\label{sec:simulation_results}
In this section, feedforward parametrization \eqref{eq:ffw_parametrization} is validated on an example dynamic system that is contained in this parametrization. It is shown that it outperforms the feedforward class of rational transfer functions and that the non-uniqueness in the parametrization is resolved by the orthogonal projection-based regularizer \eqref{eq:orthogonal_regularizer_OE_SK}.
\subsection{Example System}
The dynamic system $\mathcal{J}: f(k) \rightarrow y(k)$ is given by a two-mass-spring-damper system, see Fig. \ref{fig:example_system}, with a nonlinear damper $d_{NL}$ connecting $m_1$ to the fixed world. The discrete-time input-output behaviour from $f$ to $y$, i.e., the collocated mass, is governed by
\begin{gather}
		\sum_{i=0}^2 b_i \delta^i f(k) = \sum_{i=0}^4 a_i \delta^i y(k) + \sum_{i=0}^2 b_i \delta^i d_{NL}(\delta y(k)), 	\label{eq:example_system}
 \\
		\begin{aligned}
		b_0 &= k_2 & b_1 &= d_2 & b_2 &= m_2 & a_0 &= k_1 k_2 & a_1 &= d_2 k_1 \nonumber
		\end{aligned}
		\\
		\begin{aligned}
		 a_2 &= m_2 k_1 + k_1 m_1 - k_2 m_2 & a_3 &= m_1 d_2 & a_4 &= m_1 m_2 \nonumber
		\end{aligned}
\end{gather}
which is encapsulated by feedforward parametrization \eqref{eq:ffw_parametrization} with $N_b = 2$, $N_a = 5$, $g_i = \textrm{Id}(\cdot)$, $\psi_i = \delta^i$ and $g_\phi(r(k)) = \sum_{i=0}^2 b_i \delta^i d_{NL}(\delta r(k))$ up to approximation capabilities of $g_\phi$. The nonlinear damper represents Stribeck-like friction characteristics often found in stage systems for lithographic inspections tools, for which a simple model is given by
\begin{equation}
	d_{NL}(\delta y(k)) = c_1 \delta y(k) + \frac{c_2 - c_1}{\cosh\left(\alpha \delta y(k) \right)} \delta y(k),
\end{equation}
which is visualized in Fig. \ref{fig:SmoothStribeck}. The system parameters are given by $m_1 = 1$, $m_2 = 2$, $k_1 = 1$, $k_2 = 15000$, $d_2 = 50$, $c_1 = 1$, $c_2 = 20$, $\alpha=20$, representing a stiff connection between $m_1$ and $m_2$, resulting in a high-frequency flexible mode.
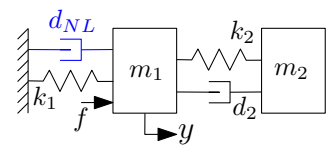
\begin{figure}[b]
\centering
\input{Linear_zeros.tex}
\caption{Two-mass-damper-spring system with nonlinear Stribeck-like friction characteristics $d_{NL}$.}
\label{fig:example_system}
\end{figure}
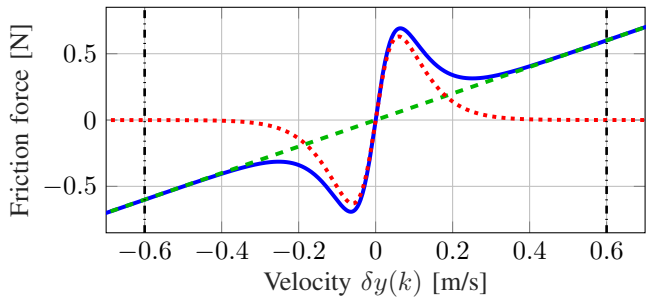
\begin{figure}[t]
	\input{SmoothStribeck.tikz}
	\caption{Stribeck-like friction curve $d_{NL}(\delta y(k))$ (\protect \drawlinelegend{mycolor1}) of example system in Fig. \ref{fig:example_system} with $c_1 = 1$, $c_2 = 20$, $\alpha = 20$, consisting of a linear (\protect \drawlinelegend{mycolor2,dashed}) and nonlinear (\protect \drawlinelegend{mycolor3,dotted}) contribution.}
	\label{fig:SmoothStribeck}
\end{figure}

For this system, a dataset of 9 references is generated combined with the optimal input $\hat{f}$ for each reference. 
\subsection{Performance Increase over Rational Basis Functions}
\label{subsec:performance_increase}
\begin{figure}[b]
	\input{ref10_f.tikz}
	\caption{The feedforward signal generated by the parallel parametrization $\mathcal{F}_{\theta,\phi}$ (\protect \drawlinelegend{mycolor2}) is able to capture the optimal input $\hat{f}$ (\protect \drawlinelegend{black, dashed}) for which $e = 0$ up to approximation capabilities, resulting in $\norm{e}{2}^2 = 0.0104$. In contrast, the feedforward signal generated by a rational transfer function (\protect \drawlinelegend{mycolor1}) is not able to correctly capture the nonlinear effects, resulting in $\norm{e}{2}^2 = 5.925$ m$^2$
.}
	\label{fig:feedforward_validation}
\end{figure}
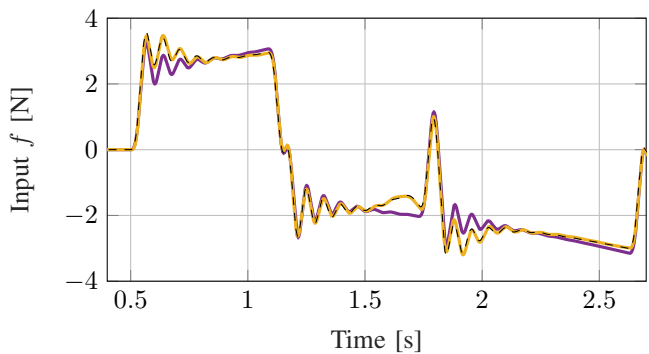

Consider the following feedforward parametrizations.
\begin{enumerate}
	\item A linear model $\mathcal{M}_\theta$ in \eqref{eq:model_class} with $g_i = \textrm{Id}$, $\psi_i(r(k)) = q^{-i+1} r(k)$ and $N_a = 10$, $N_b = 9$, corresponding to a $10^{th}$ order rational transfer function. Note that this is an overparametrization of the linear part of Fig. \ref{fig:example_system}.
	\item A parallel parametrization $\mathcal{F}_{\theta,\phi}$ in \eqref{eq:ffw_parametrization} with $g_i = \textrm{Id}$, $\psi_i(r(k)) = q^{-i+1} r(k)$ and $N_a = 5$, $N_b = 2$, and $g_\phi$ with $L=3$, $N_0 = 5$, $N_1 = 10$, $N_2 = 10$, $N_3 = 1$, i.e., the last 5 reference samples as input, 2 hidden layers and 1 output layer, with 10 neurons in each hidden layer. 
	Note that this parametrization is able to capture the dynamics up to the approximation capabilities of $g_\phi$.
\end{enumerate}
Parametrization 1) is optimized according to criterion $J_{OE}$ in \eqref{eq:J_OE} through SK-iterations, see Algorithm \ref{proc:SK_iterations}, whereas parametrization 2) is optimized with orthogonal projection-based cost function, see Algorithm \ref{proc:SK_iterations_with_OP_regularizer}. Fig. \ref{fig:feedforward_validation} shows the optimal input $\hat{f}$ and the generated input $f$ of above parametrizations for a validation reference, resulting in errors $\norm{e}{2}^2 = 5.925$ m$^2$ for the rational transfer function, and $\norm{e}{2}^2 = 0.0104$ m$^2$ for $\mathcal{F}_{\theta,\phi}$. This illustrates that $\mathcal{F}_{\theta,\phi}$ is able to effectively capture the effect of the nonlinear damper $d_{NL}(\delta y(k))$, resulting in improved performance.

\subsection{Uniqueness through Regularization}
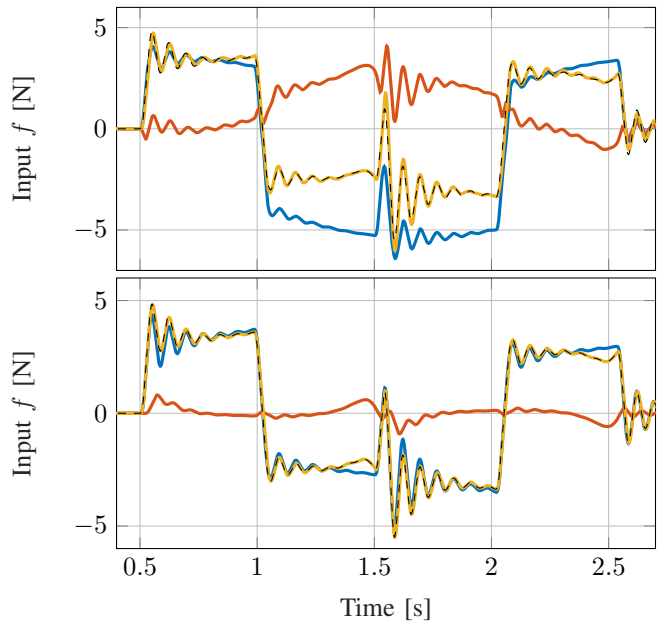
\begin{figure}[t]
	\input{ref1_non_unique_vs_unique.tikz}
	\caption{Physics-guided parametrization $\mathcal{F}_{\theta,\phi}$ trained with $J_{OE}^j$ (upper) and trained with $J_{OE,P}^j$ (lower), with optimal $\hat{f}$ (\protect \drawlinelegend{black,dashed}) and generated $f$ (\protect \drawlinelegend{mycolor3}) consisting of the sum of model-based and approximator components $f_\mathcal{M}$ (\protect \drawlinelegend{mycolor1}) and $f_\mathcal{C}$ (\protect \drawlinelegend{mycolor2}). Since $\mathcal{F}_{\theta,\phi}$ is overparametrized, there is a manifold of optimal $\theta,\phi$ for which $f_\mathcal{M} + f_\mathcal{C} = f$ equals the optimal feedforward $\hat{f}$, possibly resulting in equal but opposing contributions (upper) of $f_\mathcal{M}$ and $\mathcal{f}_\mathcal{C}$ (upper). The orthogonal projection-based regularizer removes this manifold, resulting in complementary $f_\mathcal{M}$ and $f_\mathcal{C}$ while still obtaining $f = \hat{f}$ (lower).}
	\label{fig:non_unique_vs_unique_feedforward}
\end{figure}

Consider parametrization 2) of Section \ref{subsec:performance_increase}, 
where the optimum $\theta^*, \phi^* \arg \min_{\theta,\phi} J_{OE}^j$ is non-unique, see Corollary \ref{cor:singular_subspace}. 
This non-uniqueness allows for exchanging content between $\theta$ and $\phi$ through the singular vectors $v_\theta, v_\phi$, resulting in equal but opposing contributions of $f_\mathcal{M}$ and $f_\mathcal{C}$ without changing the total feedforward $f$, see Fig. \ref{fig:non_unique_vs_unique_feedforward} (upper). In contrast, $J_{OE,P}^j$ with orthogonal projection-based regularizer \eqref{eq:orthogonal_regularizer_OE_SK} promotes uniqueness of $\theta$, see Corollary \ref{cor:unique_theta}, such that all effects that can be captured by $\mathcal{M}_\theta$, are encapsulated in $\theta$, see Fig. \ref{fig:non_unique_vs_unique_feedforward} (lower).

%% file: Linear_zeros.tex
\tikzstyle{ipe stylesheet} = [
  ipe import,
  even odd rule,
  line join=round,
  line cap=butt,
  ipe pen normal/.style={line width=0.4},
  ipe pen heavier/.style={line width=0.8},
  ipe pen fat/.style={line width=1.2},
  ipe pen ultrafat/.style={line width=2},
  ipe pen normal,
  ipe mark normal/.style={ipe mark scale=3},
  ipe mark large/.style={ipe mark scale=5},
  ipe mark small/.style={ipe mark scale=2},
  ipe mark tiny/.style={ipe mark scale=1.1},
  ipe mark normal,
  /pgf/arrow keys/.cd,
  ipe arrow normal/.style={scale=7},
  ipe arrow large/.style={scale=10},
  ipe arrow small/.style={scale=5},
  ipe arrow tiny/.style={scale=3},
  ipe arrow normal,
  /tikz/.cd,
  ipe arrows, 
  <->/.tip = ipe normal,
  ipe dash normal/.style={dash pattern=},
  ipe dash dotted/.style={dash pattern=on 1bp off 3bp},
  ipe dash dashed/.style={dash pattern=on 4bp off 4bp},
  ipe dash dash dotted/.style={dash pattern=on 4bp off 2bp on 1bp off 2bp},
  ipe dash dash dot dotted/.style={dash pattern=on 4bp off 2bp on 1bp off 2bp on 1bp off 2bp},
  ipe dash normal,
  ipe node/.append style={font=\normalsize},
  ipe stretch normal/.style={ipe node stretch=1},
  ipe stretch normal,
  ipe opacity 10/.style={opacity=0.1},
  ipe opacity 30/.style={opacity=0.3},
  ipe opacity 50/.style={opacity=0.5},
  ipe opacity 75/.style={opacity=0.75},
  ipe opacity opaque/.style={opacity=1},
  ipe opacity opaque,
]
\begin{tikzpicture}[ipe stylesheet]
  \draw
    (112, 768) rectangle (136, 736);
  \node[ipe node]
     at (118.144, 749.82) {$m_1$};
  \draw
    (80.0526, 767.907)
     -- (80, 736);
  \draw
    (76.0526, 763.9065)
     -- (80.0526, 767.9065);
  \draw
    (76.0526, 759.9065)
     -- (80.0526, 763.9065);
  \draw
    (76.0526, 755.9065)
     -- (80.0526, 759.9065);
  \draw
    (76.0526, 751.9065)
     -- (80.0526, 755.9065);
  \draw
    (76.0526, 747.9065)
     -- (80.0526, 751.9065);
  \draw
    (76.0526, 743.9065)
     -- (80.0526, 747.9065);
  \draw
    (76.0526, 739.9065)
     -- (80.0526, 743.9065);
  \draw
    (76.0526, 735.9065)
     -- (80.0526, 739.9065);
  \draw[blue]
    (80, 760)
     -- (96.053, 759.907)
     -- (92.0526, 759.907);
  \draw[blue]
    (92.0526, 763.9065)
     -- (100.0526, 763.9065)
     -- (100.0526, 755.9065)
     -- (92.0526, 755.9065)
     -- (92.0526, 755.9065);
  \draw[blue]
    (100.053, 759.907)
     -- (112, 760);
  \draw[->]
    (124.053, 735.907)
     -- (124, 728)
     -- (136, 728);
  \node[ipe node, font=\large]
     at (136.523, 726.513) {$y$};
  \draw
    (168, 768) rectangle (192, 736);
  \node[ipe node]
     at (172.705, 750.261) {$m_2$};
  \draw
    (136.032, 756.191)
     -- (141.705, 756.088)
     -- (144.032, 760.191)
     -- (148.032, 752.191)
     -- (152.032, 760.191)
     -- (156.032, 752.191)
     -- (160.032, 760.191)
     -- (162.207, 755.95)
     -- (168.128, 755.975);
  \draw
    (136.0741, 743.9892)
     -- (152.1271, 743.8962)
     -- (148.1267, 743.8962);
  \draw
    (148.1267, 747.8957)
     -- (156.1267, 747.8957)
     -- (156.1267, 739.8957)
     -- (148.1267, 739.8957)
     -- (148.1267, 739.8957);
  \draw
    (156.1271, 743.8962)
     -- (168.0741, 743.9892);
  \node[ipe node]
     at (97.028, 730.972) {$f$};
  \draw
    (80.032, 748.191)
     -- (85.705, 748.088)
     -- (88.032, 752.191)
     -- (92.032, 744.191)
     -- (96.032, 752.191)
     -- (100.032, 744.191)
     -- (104.032, 752.191)
     -- (106.207, 747.95)
     -- (112.128, 747.975);
  \node[ipe node, text=blue]
     at (87.73, 768.217) {$d_{NL}$};
  \node[ipe node]
     at (157.024, 735.357) {$d_2$};
  \node[ipe node]
     at (81.68, 739.048) {$k_1$};
  \node[ipe node]
     at (156, 764) {$k_2$};
  \draw[->]
    (100, 740)
     -- (112, 740);
\end{tikzpicture}

%% file: SmoothStribeck.tikz
%
\definecolor{mycolor1}{rgb}{0.00000,0.0,1.00000}%
\definecolor{mycolor2}{rgb}{0.00000,0.72202,0.00000}%
\definecolor{mycolor3}{rgb}{1.00000,0.0000,0.00000}%
\definecolor{mycolor4}{rgb}{0.00000,0.0000,0.00000}%

\begin{tikzpicture}

\begin{axis}[%
width=7.1650cm,
height=3cm,
scale only axis,
xmin=-.7,
xmax=.7,
xlabel style={font=\color{white!15!black},yshift=.15cm},
xlabel={Velocity $\delta y(k)$ [m/s]},
ymin=-.85,
ymax=.85,
ylabel style={font=\color{white!15!black}, yshift=-.1cm},
ylabel={Friction force [N]},
axis background/.style={fill=white},
xmajorgrids,
xminorgrids,
ymajorgrids,
yminorgrids
]
\addplot [color=mycolor1, line width=1.5pt, forget plot]
  table[]{SmoothStribeck-1.tsv};
\addplot [color=mycolor2, dashed, line width=1.5pt, forget plot]
  table[]{SmoothStribeck-2.tsv};
\addplot [color=mycolor3, dotted, line width=1.5pt, forget plot]
  table[]{SmoothStribeck-3.tsv};
\addplot [color=mycolor4, dashdotted, line width=1pt, forget plot]
  table[]{SmoothStribeck-4.tsv};
\addplot [color=mycolor4, dashdotted, line width=1pt, forget plot]
  table[]{SmoothStribeck-5.tsv};
\end{axis}
\end{tikzpicture}%

%% file: ref10_f.tikz
%

\definecolor{mycolor2}{rgb}{0.92900,0.69400,0.12500}%
\definecolor{mycolor1}{rgb}{0.49400,0.18400,0.55600}%

\begin{tikzpicture}

\begin{axis}[%
width=7.1650cm,
height=3.5cm,
scale only axis,
at={(0.758in,1.8in)},
xmin=.4,
xmax=2.7,
xlabel style={font=\color{white!15!black}},
xlabel={Time [s]},
ymin=-4,
ymax=4,
ylabel style={font=\color{white!15!black},yshift=-.1cm},
ylabel={Input $f$ [N]},
axis background/.style={fill=white},
xmajorgrids,
ymajorgrids,
ymajorgrids,
yminorgrids
]
\addplot [color=mycolor1, line width=1.2pt, forget plot]
  table[]{ref10_f-1.tsv};
\addplot [color=mycolor2, line width=1.2pt, forget plot]
  table[]{ref10_f-2.tsv};
\addplot [color=black, dashed, forget plot]
  table[]{ref10_f-3.tsv};
\end{axis}
\end{tikzpicture}%

%% file: ref1_non_unique_vs_unique.tikz
%
\definecolor{mycolor1}{rgb}{0.00000,0.44700,0.74100}%
\definecolor{mycolor2}{rgb}{0.85000,0.32500,0.09800}%
\definecolor{mycolor3}{rgb}{0.92900,0.69400,0.12500}%
\newcommand{\xminn}{.4}
\newcommand{\xmaxx}{2.7}
\newcommand{\yminn}{-6}
\newcommand{\ymaxx}{6}
\newcommand{\heightt}{3.5 cm}
\newcommand{\linewidthh}{1.2 pt}
\begin{tikzpicture}

\begin{axis}[%
width=7.1650cm,
height=3.6cm,
at={(0cm,0cm)},
scale only axis,
xmin=\xminn,
xmax=\xmaxx,
ymin=\yminn,
ymax=\ymaxx,
xlabel style={font=\color{white!15!black}},
xlabel={Time [s]},
ylabel style={font=\color{white!15!black}},
ylabel={Input $f$ [N]},
axis background/.style={fill=white},
xmajorgrids,
xminorgrids,
ymajorgrids,
yminorgrids,
]
\addplot [color=mycolor1, line width=\linewidthh, forget plot]
  table[]{ref1_unique_f-1.tsv};
\addplot [color=mycolor2, line width=\linewidthh, forget plot]
  table[]{ref1_unique_f-2.tsv};
\addplot [color=mycolor3, line width=\linewidthh, forget plot]
  table[]{ref1_unique_f-3.tsv};
\addplot [color=black, dashed, forget plot]
  table[]{ref1_unique_f-4.tsv};
\end{axis}

\begin{axis}[%
width=7.1650cm,
height=3.5cm,
scale only axis,
at={(0cm,\heightt + .2 cm)},
xmin=\xminn,
xmax=\xmaxx,
ymin=-7,
ymax=\ymaxx,
ylabel style={font=\color{white!15!black}},
ylabel={Input $f$ [N]},
axis background/.style={fill=white},
xmajorgrids,
xminorgrids,
ymajorgrids,
yminorgrids,
xlabel style={},
xticklabels={,,},
]
\addplot [color=mycolor1, line width=\linewidthh, forget plot]
  table[]{ref1_non_unique_f-1.tsv};
\addplot [color=mycolor2, line width=\linewidthh, forget plot]
  table[]{ref1_non_unique_f-2.tsv};
\addplot [color=mycolor3, line width=\linewidthh, forget plot]
  table[]{ref1_non_unique_f-3.tsv};
\addplot [color=black, dashed, forget plot]
  table[]{ref1_non_unique_f-4.tsv};
\end{axis}
\end{tikzpicture}%

%% file: Sections/Conclusion.tex
\section{Conclusion}
\label{sec:conclusion}
This paper has developed a feedforward control framework that enables superior performance over model-based feedforward control, while maintaining interpretability and task flexibility. The feedforward controller is parametrized as a parallel combination of a physics-based model and neural network, with shared autoregressive dynamics, exactly encapsulating a class of nonlinear systems with linear zero dynamics. The physics-based model and neural network are optimized simultaneously according to an output-error criterion using SK-iterations. At each SK-iteration, complementarity of the physics-based model and neural network is promoted through an iteration-dependent orthogonal projection-based regularizer. This regularizer penalizes the output of the neural network in the subspace of the model, resulting in interpretable model coefficients. The superior performance of the framework over a rational feedforward parametrization is validated on a two-mass-damper-spring system with nonlinear friction characteristics. %
\textcolor{red}{\ifthenelse{\boolean{shorten}}{\sout{Future work could focus on instrumental-variable-based solvers for output error minimization %
as well as extending the feedforward parametrization to nonlinear zero dynamics.}}{}}

%% file: Sections/Appendix.tex
\appendix
\section*{Proof of Lemma \ref{lem:J_EE_lifted}}
\begin{proof}
	The proof follows by vectorizing the signals in \eqref{eq:J_EE} over time. 
\end{proof}

\section*{Proof of Lemma \ref{lem:pinv_solution}}
\begin{proof}
	The pseudoinverse is a least-squares solution of $J_{OE}^1$ \cite{lay2003linear}. The non-uniqueness follows from the definition of the kernel of a matrix.
\end{proof}

\section*{Proof of Lemma \ref{lem:modes_of_non_uniqueness}}
\begin{proof}
	If: P1: the vector $\begin{bmatrix} 0 & v_\phi^\T \end{bmatrix}^\T \neq 0$ is in $\textnormal{ker} \begin{bmatrix}M & H(\rl)^\T \end{bmatrix}$ by virtue of $H(\rl)^\T v_\phi = 0$. 
	
	P2: there exists $u \in \mathbb{R}^{N_\phi}$ such that $M_i = H(\rl)^\T u$ \cite{lay2003linear}. Define $v = \begin{bmatrix} \epsilon_i^T &  -u^\T \end{bmatrix}^\T \neq 0$ with $\epsilon_i \in \mathbb{R}^{N_\theta}$ such that its $i^{th}$ entry equals 1, and 0 otherwise. Then $\begin{bmatrix} M & H(\rl)^\T \end{bmatrix} v = 0$.
	
	Only if: $M$ has full rank by Assumption \eqref{ass:M_full_rank}, such that above cases are the only possibilities.
\end{proof}

\section*{Proof of Theorem \ref{th:J_EE_lifted_decoupled}}
\begin{proof}
	The proof is based on orthonormality of $U_1$ and $U_2$. By Lemma \ref{lem:SVD}, 
	\begin{align}
		\fhl &= U_1 U_1^\T \fhl + U_2 U_2^\T \fhl \label{eq:fhat_decomp} \\
		H(\rl)^\T \phi^\T &= U_1 U_1^\T H(\rl)^\T \phi^\T + U_2 U_2^\T H(\rl)^\T \phi^\T. \label{eq:approx_decomp}
	\end{align}
	Combining \eqref{eq:SVD_model_resp}, \eqref{eq:fhat_decomp} and \eqref{eq:approx_decomp}, $J_{EE}$ can be written as
	{\small
	\begin{equation*}
		\norm{(U_1 U_1^\T + U_2 U_2^\T) \fhl - U_1 \Sigma V^\T \theta - (U_1 U_1^\T + U_2 U_2^\T)H(\rl)^\T \phi^\T}{2}^2.
	\end{equation*}
	}
	Since $U_1^\T U_2 = 0$, the cross-products above cancel, such that
	\begin{equation}
	\begin{aligned}
		J_{OE}^1 = &\norm{U_1 \left( U_1^\T \fhl - \Sigma V^\T \theta - U_1^\T H(\rl)^\T \phi^\T \right)}{2}^2 \\
		 &+\norm{U_2 \left(U_2^\T \fhl - U_2 H(\rl)^\T \phi^\T \right)}{2}^2,
	\end{aligned}
	\label{eq:norm_split}
	\end{equation}
	where $U_i$, $i=1,2$ are unimodular, such that $\norm{U_i(\cdot)}{2} = \norm{\cdot}{2}$. Thus, \eqref{eq:norm_split} can be equivalently written as \eqref{eq:J_EE_lifted_decoupled}, completing the proof.
\end{proof}

\section*{Proof of Corollary \ref{cor:singular_subspace}}
\begin{proof}
	As $v$ satisfies $\begin{bmatrix} M & H(\rl)^\T \end{bmatrix} v = 0$, and \eqref{eq:J_EE_lifted} is equivalent to \eqref{eq:J_EE_lifted_decoupled} by Theorem \ref{th:J_EE_lifted_decoupled}, then $v$ should satisfy \eqref{eq:matrix_subspace}. Since $\Sigma V^\T \in \mathbb{R}^{N_\theta \times N_\theta}$ is full rank by Assumption \eqref{ass:M_full_rank}, \eqref{eq:singular_subspace_expression} follows from the first rows of \eqref{eq:matrix_subspace}.
\end{proof}

\section*{Proof of Theorem \ref{th:J_EE_lifted_OP_decouped}}
\begin{proof}
	The proof follows along the same lines as the proof to Theorem \ref{th:J_EE_lifted_decoupled}, i.e., $J_{OE,P}^1$ can be written as
	\begin{align}
		J_{OE,P}^1 =& \norm{U_1 \left( U_1^\T \fhl - \Sigma V^\T \theta - U_1^\T H(\rl)^\T \phi^\T \right)}{2}^2 \nonumber \\
		 &+\norm{U_2 \left(U_2^\T \fhl - U_2 H(\rl)^\T \phi^\T \right)}{2}^2 \label{eq:norm_split_OP} \\ 
		 &+ \lambda \norm{(\Sigma V^\T)^{-1} U_1^\T H(\rl)^\T \phi^\T}{2}^2, \nonumber
	\end{align}
	which is equivalent to \eqref{eq:J_EE_lifted_OP_decoupled}, completing the proof.
\end{proof}

\section*{Proof of Corollary \ref{cor:unique_theta}}
\begin{proof}
Denote the cost corresponding to $x^*$ by 
\begin{equation}
J_{x^*} = \norm{
	\begin{bmatrix}
		U_1^\T \fhl 
		\\ U_2^\T \fhl 
	\end{bmatrix}
	-
	\begin{bmatrix}
		\Sigma V^\T & U_1^\T H(\rl)^\T \\
		0 & U_2^\T H(\rl)^\T
	\end{bmatrix}
	x^*
	}{2}^2.
\end{equation}
	Consider $J_{OE}^1$ in \eqref{eq:J_EE_lifted_decoupled}. For any $v \in \textnormal{ker} \begin{bmatrix}M & H(\rl)^\T \end{bmatrix}$ it holds that
	\begin{align}
	J_{OE}^1(x^*+v) = &\norm{
	\begin{bmatrix}
		U_1^\T \fhl 
		\\ U_2^\T \fhl 
	\end{bmatrix}
	-
	\begin{bmatrix}
		\Sigma V^\T & U_1^\T H(\rl)^\T \\
		0 & U_2^\T H(\rl)^\T
	\end{bmatrix}
	(x^* + v)
	}{2}^2 \nonumber
	\\
	=& J_{x^*},
	\end{align}
	since $v_1 = \begin{bmatrix} v_\theta^\T & v_\phi^\T \end{bmatrix}^\T \in \textnormal{ker} \begin{bmatrix}M & H(\rl)^\T \end{bmatrix}$. Hence
	\begin{equation}
		J_{OE}^1(x^* + v_1) = J_{OE}^1(x^* + v_2).
	\end{equation}		
	 In contrast, for $J_{OE,P}^1$ in \eqref{eq:J_EE_lifted_OP_decoupled} and any $v \in \textnormal{ker} \begin{bmatrix}M & H(\rl)^\T \end{bmatrix}$ it holds that
	\begin{align}
		J_{OE,P}^1(x^* + v) =& \norm{
	\begin{bmatrix}
		U_1^\T \fhl 
		\\ U_2^\T \fhl 
	\end{bmatrix}
	-
	\begin{bmatrix}
		\Sigma V^\T & U_1^\T H(\rl)^\T \\
		0 & U_2^\T H(\rl)^\T
	\end{bmatrix}
	x^* 
	}{2}^2 \nonumber \\
	&+ \norm{\sqrt{\lambda}(\Sigma V^\T)^{-1} U_1^\T H(\rl)^\T v_\phi}{2}^2 \\
	=& J_{x^*} + \norm{\sqrt{\lambda} v_\theta}{2}^2, \nonumber
	\end{align}
	in which the last identity holds by \eqref{eq:singular_subspace_expression}. Then
	\begin{equation}
		J_{OE,P}^1(x^* + v_1) < J_{OE,P}^1(x^* + v_2),
	\end{equation}
	completing the proof.
\end{proof}